\documentclass[newstyle,twocolumn,proceedings]{rmaa}
\title{Hydrodynamic Modelling of Accretion Flows}
\author{
J.~R.~Murray,\altaffilmark{1} 
M.~.R.~Truss,\altaffilmark{2}
S.~B.~Foulkes,\altaffilmark{3}
C.~A.~Haswell\altaffilmark{3}
and K.~Manson\altaffilmark{4}}
\altaffiltext{1}{Swinburne University of Technology, Australia.}
\altaffiltext{2}{University of St Andrews, UK.}
\altaffiltext{3}{Open University, Milton Keynes, UK.}
\altaffiltext{4}{Australian National University, Australia.}

\fulladdresses{
\item James R. Murray: Centre for Astrophysics \& Supercomputing,
Swinburne University of Technology, Hawthorn, VIC 3122, Australia
(\email{jmurray@swin.edu.au})
\item Carole Haswell \& Steven Foulkes: Department of Physics \& Astronomy,
The Open University,  Walton Hall, Milton Keynes MK7 6AA, U.K.
\item Kathy Manson: Schol of Mathematical Sciences,
The Australian National University,  ACT 0200, Australia
\item Michael Truss: Department of Physics \& Astronomy,
The University of St Andrews,   North Haugh, St Andrews KY16 9SS, Scotland
}

\shortauthor{J.~R. Murray et al.}
\shorttitle{Hydrodynamic Modelling of Accretion Flows}
\abstract{In the proceedings of this, and of several recent close binary conferences, there have been 
several contributions describing smoothed particle hydrodynamics simulations of accretion disks. It is
apposite therefore to review the numerical scheme itself with emphasis on its advantages for disk modelling, and 
the methods used for modelling viscous processes.
}
\addkeyword{accretion, accretion disks}
\addkeyword{binaries:close}
\addkeyword{line:profiles}
\addkeyword{x-rays:binaries}
\addkeyword{hydrodynamics}
\addkeyword{methods:numerical}

\begin{document}
\maketitle
\section{Smoothed particle hydrodynamics}
Smoothed particle hydrodynamics or SPH is a numerical scheme for modelling the motion of gases, 
fluids and solids. SPH is a particle technique. In other words, the fluid is represented by a set of particles. The
motion of the particles represents the fluid motion, and the interactions of the particles represent the fluid forces.
Each particle has a mass $m$ which remains fixed, and a length scale $h$ over which properties of the fluid are interpolated.
If a simulation were equated to an astronomical image, then $h$ would be the width of the point spread function.

Interpolation is the key to SPH. Fluid properties, at any location, can be determined by interpolation of the particles' properties. 
For example the density at any point in the fluid can be determined by interpolating the particle masses. 
The fluid velocity field can
be determined by interpolating the particles' velocities.

By way of example, we demonstrate how the density could be estimated at a general point ${\bf r}$ in the fluid, by using a 
simple weighted sum of the particle masses. 
\begin{equation}
\rho({\bf r}) = \sum_{i=1}^N m_{\rm i} \, K\,exp (-({\bf r}-{\bf r}_{\rm i})^2/h^2.
\end{equation}
We have chosen a Gaussian interpolation function (with normalisation factor $K$) to re-emphasise the link with point spread functions.
(In practice Gaussian interpolation is never used with SPH as it would require summation over all $N$ particles in the simulation 
every time an interpolation was called for).

Using the principle of interpolation as illustrated briefly above, and by dint of heavy mathematics, the fluid equations can be 
rewritten as a set of equations for the motion of a set of particles. These SPH equations can then be solved using your favourite algorithm
for integrating differential equations (Euler, leap-frog, Runge-Kutta, etc.). There are two important points to make. Firstly,
just as the fluid equations can be rewritten in a number of ways, there are many possible SPH equivalents to the fluid equations. A good
set of SPH equivalents are those that allow conservation of fundamental fluid properties and that are conducive to a stable numerical solution.
The second point is that, in the limit of infinitely many particles, and each particle having many neighbours, properly formed SPH
equations simply reduce to the original fluid equations. In this very important sense then, SPH is a computational fluid dynamics method, and
it should not be confused with so-called sticky particle schemes.

The basic SPH methodology is reviewed in Monaghan (1992), however many algorithmic developments have taken place
since then, to the point where an up to date general review article is now sorely needed. In this article we will focus
on algorithmic advances that are relevant to those modelling accretion disks. 

\section{Modelling viscous disks}
Smoothed particle hydrodynamics has a reputation for being overly viscous, as the technique relies upon an artificial 
viscosity term in the equations of motion to resolve hydrodynamic shocks. This would make it difficult to model close binary
accretion disks which feature both an anomalously large viscosity, and shocks. However, the reputation is ill-deserved, as
the amount and form of the viscosity can be controlled in a number of ways.

Any computational fluid dynamics algorithm must incorporate a scheme to account for shocks. These are regions of the flow in
which the basic hydrodynamic variables are discontinuous on a length scale comparable to the mean free path of the 
individual particles comprising the gas or fluid. For any realistic calculation, the mean free path, $\lambda$, will be several 
orders of magnitude smaller than the
resolution, $l$, and so they can never be modelled completely faithfully. The best a numerical scheme can hope for is to be able to 
reliably predict the location, strength and speed of propagation of any discontinuity in the fluid variables by 
recording a similar discontinuity in the 
numerical variables. This is a sore test indeed as errors in computational methods always contrive to smear out discontinuities
Numerical artifice must be resorted to in order to maintain the steep gradient in fluid properties over the shortest length scales 
resolvable by the simulation. 

As described in Monaghan (1992), SPH makes use of an artificial viscosity term to resolve shocks.  Originally only turned on for compression
(i.e. when particles are approaching one another), Meglicki et al. (1995) adapted the standard artifical viscosity to provide a measurable
shear viscosity in discs by removing the requirement that the fluid be in compression. They then obtained a mathematical expression for the 
dissipation produced by such a term. In Murray (1996) the expression of Meglicki et al. was compared to the results of ring spreading tests
and found to be generally accurate to within 10 \%. This paper also describes how SPH with this modified artifical viscosity term can be used
to model the behaviour of a disc with a given Shakura-Sunyaev viscosity. Using artificial viscosity to simulate a large scale 
turbulent viscosity does not imply that hydrodynamic shocks cannot be modelled. Foulkes et al. (2004) describes a number of simulations in
which the shock structure is clearly well resolved. 

Being able to control the viscosity has proved very useful when modelling the outbursts of disks in close binaries. The approach was 
originally demonstrated in Truss et al. (2000), and has subsequently been used to explain the outbursts of a number of different systems
(Truss, Murray \& Wynn, 2001; Truss et al. 2002, Truss, Wynn \& Wheatley 2004). The biggest success of these papers has been the clear
demonstration that a local thermal instability can give rise to disc wide changes in structure.

Other workers have taken different approaches. Flebbe et al. (1994) wrote down the SPH equivalent of a general Navier Stokes viscosity.
This gave them the ability to vary the ratios of the shear and bulk viscosity coefficients, which are held in fixed ratio if the standard
artificial viscosity is used. Flebbe et al.'s approach has been incorporated into a number of simulations by Kunze and co-workers. In 
Kunze, Speith \& Hessman (2001), the stream-disk impact in cataclysmic variables was studied in considerable detail.

In all the above work, disks with a substantial viscosity are being modelled. The question remains, can SPH be used if shear viscosity is
not the primary force driving disc evolution? Most assuredly yes.  Maddison, Humble \& Murray (2003) showed that SPH could be used to
simulate protoplanetary discs in which gas-dust drag rather than gas viscosity drives evolution. In order to do so, two further 
advances were made use of. Balsara (1995) developed a modification of the artificial viscosity term that approximately sets the dissipation
to zero in regions of pure shear, whilst leaving it unaffected in regions of compression. This is done by obtaining SPH estimates of the
curl and divergence of the velocity field, and then forming a coefficient for the artificial viscosity term,
\begin{equation}
K_{\rm v} = \frac { |{\bf \nabla \cdot v}|}{ |{\bf \nabla \cdot v}| +  |{\bf \nabla \times v}|}.
\end{equation}
Such a term is also used successfully by cosmologists who find the removal of angular momentum from galactic disks to be distasteful. 
Steinmetz (1996) analyzed the effectiveness of this term in quite considerable detail.  Morris \& Monaghan (1997) carried the process 
further by allowing each SPH particle to have  a time varying artificial viscosity coefficient. The coefficient becomes large when a 
particle approaches a shock front, and then decays as the particle passes into the shocked region of the flow. Although no tests of the
relative effectiveness of these two viscosity limiting techniques have been published to date, our experience is that the Balsara
modifier is of most use in disk situations.  

\section{Discussion}
One of the great advantages of the smoothed particle hydrodynamics technique is its simplicity. An SPH code is rarely more than a 
thousand lines long. This, in combination with established methods for incorporating new physics into the algorithm, makes it an ideal
tool for studying systems where the important physics is uncertain. In the case of accretion disk research, we can not yet describe with
any confidence the form of the anomalously large dissipation. We can however make use of SPH simulations to investigate various postulated
forms for disk viscosity (e.g. the viscoelastic description developed by Ogilvie, 2003). The encouraging message is thus that if there is 
viscosity in an SPH simulation of a disk, it is there specifically to model the unknown process driving accretion in close binaries. It is
without doubt possible to have shocks and a prescribed viscous dissipation term in the same simulation.
We conclude by referring the reader to presentations of SPH disk simulations at this conference by Hayasaki, Kunze, Manson, Truss and
Okazaki.

\end{document}